\documentclass{iaus}
\usepackage{graphicx}

\title[Polarization modeling with STOKES]
      {AGN polarization modeling with {\sc Stokes}}

\author[Goosmann, Gaskell \& Shoji]
       {R.~W. Goosmann$^1$, C. Martin Gaskell$^2$, \and M. Shoji$^3$}

\affiliation{
  $^1$ Astronomical Institute, Academy of Sciences, Prague, Czech Republic
       \\[\affilskip]
  $^2$ Dept. Physics \& Astronomy, Univ. of Nebraska, Lincoln, NE, USA
       \\[\affilskip]
  $^3$ Astronomy Dept., University of Texas, Austin, TX, USA
       }

\pubyear{2006}
\volume{238}  %% insert here IAU Symposium No.
\pagerange{001--999}
\date{??? and in revised form ???}
\jname{Black Holes: from Stars to Galaxies -- across the Range of Masses}
\editors{V. Karas \& G. Matt, eds.}
\begin{document}

\maketitle

\begin{abstract}
We introduce a new, publicly available Monte Carlo radiative
transfer code, {\sc Stokes}, which has been developed to model
polarization induced by scattering off free electrons and dust
grains. It can be used in a wide range of astrophysical
applications. Here, we apply it to model the polarization produced
by the equatorial obscuring and scattering tori assumed to exist in
active galactic nuclei (AGNs). We present optical/UV modeling of
dusty tori with a curved inner shape and for two different dust
types: one composition reproduces extinction properties of our
Galaxy, and the other is derived from composite quasar spectra. The
polarization spectra enable us to clearly distinguish between the
two dust compositions. The {\sc Stokes} code and its documentation
can be freely downloaded from http://www.stokes-program.info/.


\end{abstract}

\firstsection % if your document starts with a section,

\section{Introduction}

Spectropolarimetric data in the optical/UV range put important
constraints on the emission and scattering geometry of AGNs (see
e.g. \cite{antonucci2002}). For the interpretation of
spectropolarimetric data detailed modeling tools are important. Here,
we use a new, publicly available radiative transfer code, {\sc Stokes}
(\cite[Goosmann \& Gaskell\ 2007]{goosmann2007}), to model the
obscuring torus of AGNs. The code is based on the Monte-Carlo method
and allows the simulation of various emission and scattering
geometries. Polarization due to Thomson and dust (Mie-)scattering is
included. Moreover, the code computes wavelength-dependent time delays
and can thus be used to study polarization reverberation (\cite[Shoji,
Gaskell, \& Goosmann\ 2005]{shoji2005}).

\section{Modeling an obscuring torus}

We consider a torus with an elliptical cross-section and centered on a
point source. The source isotropically emits a flat continuum spectrum
between 1600~\AA \ and 8000~\AA. The torus half-opening angle is set to
$\theta_0 = 30^\circ$. The inner and outer radii of the torus are fixed at
0.25 pc and 100 pc respectively. The radial optical depth in the equatorial
plane is 750 for the V-band. The dust models (table~\ref{tab}) assume
a mixture of graphite and ``astronomical silicate'' and a grain radii
distribution $n(a) \propto a^\alpha_s$ between $a_{\rm min}$ and
$a_{\rm max}$.

\begin{table}\def~{\hphantom{0}}
  \begin{center}
    \caption{Parameterization of the dust models}
    {\tiny
    \label{tab}
    \begin{tabular}{lccccc}\hline
      Type & Graphite & Silicate & $a_{\rm min}$ & $a_{\rm max}$ &
      $\alpha_s$\\\hline
      Galactic & 62.5\% & 37.5\% & 0.005$\, \mu{\rm m}$ & 0.250$\, \mu{\rm m}$
               & $-3.5$\\
      AGN      &  85\%  &  15\%  & 0.005$\, \mu{\rm m}$ & 0.200$\, \mu{\rm m}$
               & $-2.05$\\\hline
    \end{tabular}
    }
  \end{center}
\end{table}

The ``Galactic dust'' model reproduces the interstellar extinction
for $R_{\rm V} = 3.1$ whilst the ``AGN dust'' parameterization is
obtained from quasar extinction curves derived by \cite[Gaskell
\etal\ 2004]{gaskell2004}. This latter dust type favors larger grain
sizes.

\section{Results and discussion}

\begin{figure}
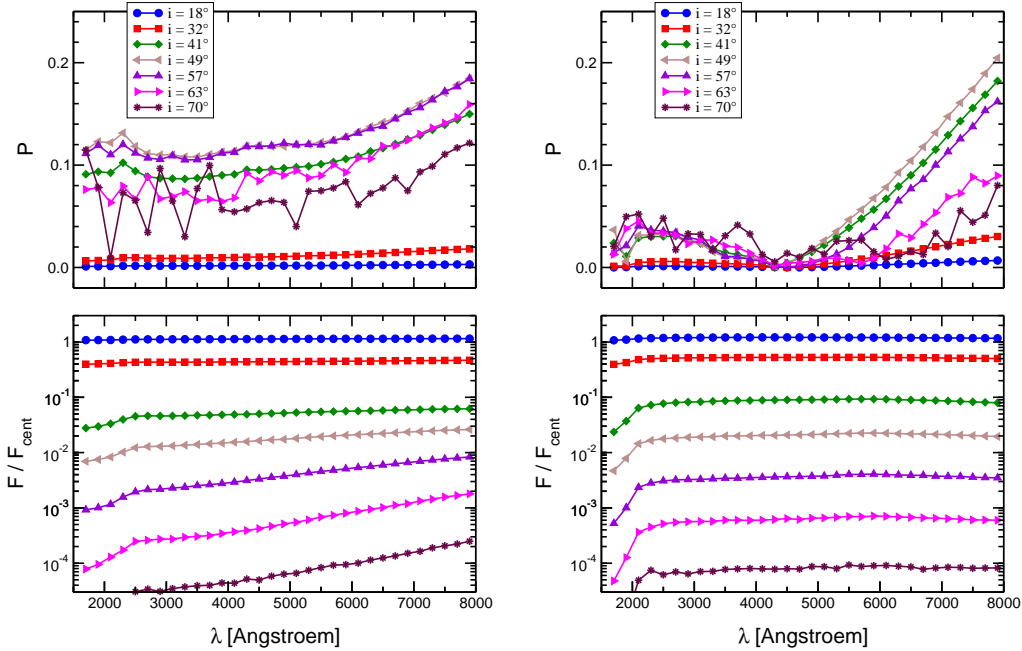

  \vskip +0.4cm
  \includegraphics[width=0.48\textwidth]{S238-goosmann-poster2-fig1.eps}
  \hfill
  \includegraphics[width=0.48\textwidth]{S238-goosmann-poster2-fig2.eps}
  \caption{Polarization (top) and flux (bottom) spectra for a
  centrally-illuminated torus filled with ``Galactic dust'' (left) or
  ``AGN dust'' (right). The flux spectra are normalized to the
  illuminating flux and the inclination angle $i$ is measured from the
  symmetry axis.}
  \label{fig}
\end{figure}

The angular distributions of the polarization and flux spectra are
shown in figure~\ref{fig}. In a face-on view, the central source is
seen directly and the obtained polarization is low. At obscured
inclinations $i>\theta_0$, the scattering properties of both dust
types lead to different results. For ``AGN dust'', the polarization is
lower in the UV but rises quickly toward longer wavelengths.  Detailed
spectropolarimetric observations of dust reflection thus effectively
constrain the dust composition. For both dust types the polarization
position angle is oriented perpendicularly to the projected symmetry
axis of the object.

If the torus is filled with ``AGN dust'' the total flux spectra are
wavelength-independent above 2500~\AA \ at all possible
inclinations. For the ``Galactic dust'' model the scattered flux rises
gradually toward longer wavelengths. The albedo of both dust
compositions changes significantly below 2500~\AA \ so that less
radiation is scattered in the UV.

Spectropolarimetric data from AGNs contain not only information
about the obscuring torus, but also include other scattering
regions. To obtain a more detailed picture of AGN polarization it is
therefore necessary to model several, radiatively coupled scattering
regions self-consistently. While this is beyond the scope of this
proceedings note, {\sc Stokes} is capable of solving such problems.

\end{document}